\documentclass[12pt,preprint]{aastex}

\usepackage{color}
\usepackage{ulem}

\newcommand{\vt}[1]{\mbox{\boldmath{$#1$}}}
\slugcomment{}

\shorttitle{Constraints on the Evolution of the Primordial Magnetic Field}
\shortauthors{Yamazaki et al.}
\begin{document}
\title{Constraints on the Evolution of the Primordial Magnetic Field from the Small Scale CMB Angular Anisotropy}
\author{D. G. Yamazaki\altaffilmark{1,2}, K. Ichiki\altaffilmark{2}, T. Kajino\altaffilmark{2,1} and G. J. Mathews\altaffilmark{3} }
\email{yamazaki@th.nao.ac.jp}
\altaffiltext{1}{Department of Astronomy, Graduate School of Science, University of Tokyo, 7-3-1 Hongo, Bunkyo-ku, Tokyo, 113-0033, Japan}
\altaffiltext{2}{Division of Theoretical Astronomy, National Astronomical Observatory Japan, 2-21-1, Osawa, Mitaka, Tokyo, 181-8588, Japan}
\altaffiltext{3}{Center for Astrophysics,
Department of Physics, University of Notre Dame, Notre Dame, IN 46556, U.S.A.}

\begin{abstract}
Recent observations  of the cosmic microwave background (CMB) have extended measured the power spectrum to higher multipoles $l\gtrsim$1000, and there appears to be possible evidence for excess power on small angular scales.
The primordial magnetic field (PMF) can strongly affect the CMB power spectrum and the formation of large scale structure.
In this paper, we calculate the CMB temperature anisotropies generated by including a power-law magnetic field at the photon last scattering surface (PLSS). 
We then deduce an upper limit on the primordial magnetic field based upon our theoretical analysis of the power excess on small angular scales. 
We have  taken into account several important effects such as reionization and the modified matter sound speed in the presence of a magnetic field.
An upper limit to the field strength of $|B_\lambda|\lesssim$ 4.7 nG at the present scale of 1 Mpc is deduced.  This is obtained by comparing the calculated theoretical result including the Sunyaev-Zeldovich (SZ) effect with recent observed data on the small scale CMB anisotropies from the Wilkinson Microwave Anisotropy Probe (WMAP), the Cosmic Background Imager (CBI) and the Arcminute Cosmology Bolometer Array Receiver (ACBAR). 
We discuss several possible mechanisms for the generation and evolution of the PMF.
\end{abstract}
\keywords{cosmology: cosmic microwave background --- methods: numerical --- 
magnetic field} 
 \newcounter{one}
 \setcounter{one}{1}
 \newcounter{two}
 \setcounter{two}{2}
 \newcounter{three}
 \setcounter{three}{3}

\section{Introduction}
The possible existence of a primordial magnetic field (PMF) is an important question in modern cosmology.
The PMF could influence a variety of phenomena in the early universe (Tsagas and Maartens 2000; Grasso and Rubinstein 2001; Betschart et al. 2003; Bruni et al. 2003; Clarkson et al 2003; Giovannini 2004; Vall$\acute{\mathrm{e}}$e 2004; Challinor 2005; Kahniashvili et al. 2005; Dolgov 2005; Kosowsky 2005; Gopal and Sethi 2005; Tashiro et al. 2005; Tashiro and Sugiyama 2005).
However, there is  still no firm understanding of the origin and evolution of the PMF, especially the observable magnetic field of $0.1-1.0~\mu$ G in galaxy clusters (Clarke et al 2001; Xu et al.~2006).
Recently, the origin of the PMF on such scales has been studied by a number of authors (Quashnock 1989; Boyanovsky et al.~2002; Bamba and Yokoyama 2004; Berezhiani and Dolgov 2004; Hanayama et al.~2005; Takahashi et al.~2005; Ichiki et al.~2006). There is, however, a possible discrepancy between theory and observation on the scale of galactic clusters.
Temperature and polarization anisotropies in the CMB
 provide very precise information on the physical processes in operation during the early universe.
However, the CMB data from the Wilkinson Microwave Anisotropy
Probe (WMAP; Bennett, et al. 2003), the Arcminute Cosmology
Bolometer Array Receiver (ACBAR; Kuo et al. 2004), and the Cosmic
Background Imager (CBI; Mason et al. 2003) have indicated
a potential discrepancy between the best-fit cosmological model to the WMAP data and observations at higher
multipoles $l \ge 1000$.
A straightforward extension of the fit to the WMAP data predicts a rapidly declining power spectrum 
in the large multipole range due to the finite thickness of the photon last scattering surface (PLSS).  However,
the  ACBAR and CBI  experiments indicate continued power up to $l \sim 4000$. 
This discrepancy is difficult to account for by a simple retuning of cosmological parameters.

One possible interpretation of such excess power at high multipoles
is a manifestation of the re-scattering of CMB photons by hot electrons
in clusters known as the thermal SZ effect (Sunyaev and
Zeldovich 1980). Although there
can be no doubt that some contribution from the SZ effect exists in
the observed angular power spectrum, it has not yet been established conclusively that this is the only possible
interpretation (Aghanim et al. 2001) of the small scale power. Indeed,
the best value of the matter fluctuation amplitude, $\sigma_8$ to fit
the excess power at high multipoles is near the upper end of the range of 
the values deduced by the other independent methods (Bond et al.~2005; 
Komatsu and Seljak 2002). Toward a solution of this problem, it has recently been proposed (Mathews et al.~2004) that a bump feature in the primordial
spectrum may provide a better explanation for both the CMB and matter power
spectra at small scales.

In this paper we consider another possibility.  Inhomogeneous cosmological magnetic field generated before the CMB last-scattering epoch is also a plausible mechanism to produce excess power at
high multipoles. Such a field excites an Alfven-wave mode in the baryon-photon
plasma in the early universe and induces small rotational velocity 
perturbations. Since this mode can survive on scales below those at which Silk
damping occurs during recombination (Jedamzik et al. 1998; Subramanian and Barrow
1998a), it could be a new source of the CMB anisotropies on small 
angular scales. 
Analytic expressions for the temperature and polarization angular power
spectra on rather larger angular scales ($l \le 500$) by Mack et al. (2002) based upon the thin PLSS approximation have been derived for both vector and tensor modes. Subramanian and Barrow (2002) considered the vector perturbations in the opposite limit of smaller angular scales. 
Non-gaussianity in fluctuations from the PMF has also been considered (Brown and Crittenden 2005).
A strong magnetic field in the early universe, however, may conflict with the cosmological observations currently
available. The combination of those studies and current observations
places a bound on the strength of the PMF of 
$B <1.0\ -$ 10 nG.

In order to compare the CMB anisotropy induced by the PMF with observations more precisely, we need to perform
numerical calculations of the fully linearized MHD equations along with a realistic recombination history of the universe. In particular, we first need to develop a numerical method to predict the theoretical spectrum for intermediate angular
scales, in which the analytic approximation becomes
inappropriate. Recently, efforts along this line have been done (Lewis 2004) to obtain numerical
estimates of the effects of the PMF on the CMB. 
Also Yamazaki et al. (2005a) have proven numerically that the PMF is an important physical process for higher multipoles $l$. They have constrained an upper limit to the PMF of $B < 3.9$ nG, based upon a likelihood analysis of the data (Yamazaki et al.~2005b).

The WMAP data is very precise, but it does not extend to $l>$900. The PMF, however, affects the CMB power spectrum for  higher multipoles $l$ (Yamazaki et al 2005b; Lewis 2004). 
Thus, one purpose of this paper is to place a new limit on the strength of the primordial magnetic field by comparing
 our calculated result with the WMAP, CBI and ACBAR data, the latter two of which have measured CMB anisotropies for higher multipoles $l$ than WMAP. We will also discuss the consistency between our CMB constraints and  the magnetic field in galactic clusters.
\section{Primordial Stochastic Magnetic Field}
The Photon-baryon
fluid can be assumed as single fluid since these particles interact each other very tightly before last scattering of photons.
Since the trajectory of plasma particles is bent by Lorentz forces in a
magnetic field, photons are indirectly influenced by the magnetic field
through Thomson scattering. 
We consider that the primordial magnetic field is generated from some moments during the radiation-dominated ear. In the case of the falt Friedmann-Robertson-Walker (FRW) background cosmology for the linear perturbation, since we treat the energy density of the primordial magnetic field as a first order perturbation and a stiff source for time evolution, the all back reactions from the fluid can be discarded.

The electromagnetic tensor has the usual form
\begin{eqnarray}
{F^\alpha}_\beta=
\left(
    \begin{array}{cccc}
      0    &  E_1  &  E_2  &  E_3 \\
      E_1  &  0    & -B_3  &  B_2 \\
      E_2  &  B_3  &  0    & -B_1 \\
      E_3  & -B_2  &  B_1  &  0  \\
    \end{array}
\right)~~,
\label{eq_emf_tensor}
\end{eqnarray}
where $E_i$ and $B_i$ are the electric and magnetic fields.
Here we use natural units $c = \hbar = 1$. 
The energy momentum tensor for electromagnetism is 
\begin{eqnarray}
{T^{\alpha\beta}}_{[\mathrm{EM}]}=\frac{1}{4\pi}\left(F^{\alpha\gamma}F^\beta_\gamma
-\frac{1}{4}g^{\alpha\beta}F_{\gamma\delta}F^{\gamma\delta}\right)\label{eq_ememtensor}.
\end{eqnarray}
The Maxwell stress tensor, $\sigma^{i k}$ is derived from the space-space components of the electromagnetic energy momentum tensor,  
\begin{eqnarray}
-{T^{ik}}_{[\mathrm{EM}]}=
\sigma^{ik}=  \nonumber \\
\frac{1}{a^2}\frac{1}{4\pi}\left\{E^i E^k+B^i B^k
- \frac{1}{2}\delta^{ik}(E^2+B^2)\right\}\label{eq_MST1}.
\end{eqnarray}
The conductivity of the primordial plasma is very large and the electric field is negligible, i.e.~$E\sim 0$. We call this condition ''frozen-in'' (Mack et al. 2002).
 This is a very good approximation during the epochs of interest here.
 In this way we obtain the following equations,
\begin{eqnarray}
{T^{00}}_{[\mathrm{EM}]}=\frac{B^2}{8\pi a^6} \label{eq_MST_00}~~, \\
{T^{i0}}_{[\mathrm{EM}]}={T^{0k}}_{[\mathrm{EM}]}=0 \label{eq_MST_0s} ~~,\\
-{T^{ik}}_{[\mathrm{EM}]}=\sigma^{ik}=\frac{1}{8\pi a^6}(2B^i B^k -
\delta^{ik}B^2)~~.
\label{eq_MST_ss}
\end{eqnarray}
\subsection{Two-Point Correlation Function} 
We assume that the PMF,  $\mathbf{B}_0$, deviates from a homogeneous and  isotropic distribution in a statistically random way. 
The power spectrum of fluctuations from a homogeneous and isotopic distribution can then be taken as a power-law $P(k)\propto k^{n_B} $\cite{m1} where $k$ is the wave number, and the spectral index $n_B$ can be either positive or negative.
Evaluating the two-point correlation function of the electromagnetic stress-energy tensor, we can obtain a good approximation to the the vector isotropic spectrum for $k<k_C$ (Mack et al. 2002)\footnote{This has a form similar to that used in Mack et al.~(2002) who treated two kinds of Fourier transforms of different normalizations. Here, we systematically adopt the same normalization for the Fourier transform. This removes the some of the uncertainty in the numerical calculations (Lewis 2004).}, 
\begin{eqnarray}
|\Pi^{(1)}(\mathbf{k})|^2\simeq
\frac{1}{4(2n_B+3)}\left[\frac{(2\pi)^{n_B+3}B^2_\lambda}{2\Gamma\left(\frac{n_B+3}{2}\right) k^{n_B+3}_\lambda}\right]^2\nonumber \\
\times\left({k_C}^{2n_B+3}+\frac{n_B}{n_B+3}k^{2n_B+3}\right),\label{eq:Pi}
\end{eqnarray}
where the vector mode Lorentz force, $L(\mathbf{k})$, is given by $L(\mathbf{k})=k\Pi^{(1)}(\mathbf{k})$, 
$B_\lambda=|\mathbf{B}_\lambda|$ is the magnetic strength for the comoving mean scale derived by smoothing over a Gaussian sphere for radius $\lambda$,(where $\lambda=1$ Mpc in this paper). 
In this equation $k_C$ is the cutoff wave number in the magnetic power spectrum defined by
\begin{eqnarray}
k_C\simeq \left(5.08\times 10^{-2}\times\frac{B_\lambda}{1\mbox{nG}}\right)^{2/n_B+5} \nonumber \\
\times\left(\frac{k_\lambda}{1\mbox{Mpc}^{-1}}\right)^{\frac{n_B+3}{n_B+5}}
\left(\Omega_bh^3 \right)^{1/n_B+5}~~,
\label{eq:kd}
\end{eqnarray}
where $h$ is the Hubble parameter in units of 100 km s$^{-1}$ Mpc$^{-1}$, (Mack et al. 2002; Jedamzik, et al. 1998; Subramanian and Barrow 1998a).
We obtain the explicit time dependence of the primordial magnetic stress as $\Pi^{(1)}(\eta,{\mathbf k})=\Pi^{(1)}({\mathbf k})/a^4$ since $\Pi^{(1)}({\mathbf k})$ are affected from the primordial magnetic field quadratically.
The scalar mode of the magnetic field stress tensor $\Pi^{(0)}$ is obtained via a  similar calculation (Koh and Lee 2002)
\begin{eqnarray}
|\Pi^{(0)}(\mathbf{k})|^2\simeq
\frac{1}{2(2n_B+3)}\left[\frac{(2\pi)^{n_B+3}B^2_\lambda}{2\Gamma\left(\frac{n_B+3}{2}\right) k^{n_B+3}_\lambda}\right]^2\nonumber \\
\times\left({k_C}^{2n_B+3}+\frac{n_B}{n_B+3}k^{2n_B+3}\right)~~.
\label{eq:Scalar_Pi}
\end{eqnarray}
\subsection{Ionization ratio}
In previous studies the ionization ratio $x_e$ in the early universe has been  assumed to obey  a step function in time, i.e. $x_e=1$ before the PLSS and  $x_e=0$ after the PLSS. However, more accurate theoretical results require the use of the exact ionization ratio. We have therefore carried out a numerical estimate of the ionization ratio by using the program "RECFAST" (Seager et al. 1999).
In this paper we take exact account of the  ionization ratio $x_e(\tau)$, and can now rewrite the Lorentz force as 
\begin{eqnarray}
L(\tau, \mathbf{k})=kx_e(\tau)\Pi^{(1)}(\mathbf{k})~~.
\label{eq:ionization}
\end{eqnarray}

\section{Perturbation Evolution Equations}
To obtain the scalar and vector perturbation evolution equations we write the perturbed Einstein equation as
\begin{eqnarray}
(\bar G_{\mu\nu}+\delta G_{\mu\nu})&=8\pi G
(\bar{T}_{\mu\nu}{}_{[\mathrm{FL}]}+\delta T_{\mu\nu}{}_{[\mathrm{FL}]})~~,
\label{eq_Einstein} 
\end{eqnarray}
where $T_{\mu\nu}{}_{[\mathrm{FL}]}$ is the energy-momentum tensor of a perfect fluid.
The Einstein equation can then be separated into background and perturbation equations.
The background Einstein tensor is given by\\
\begin{eqnarray}
\bar G_{\mu\nu}=
\bar R^\mu{}_\nu -\frac{1}{2}\delta^\mu{}_\nu\bar R~~, 
\label{eq_EinsteinTensorB}
\end{eqnarray}
while the perturbation tensor is
\begin{eqnarray}
\delta G_{\mu\nu}=
\delta R^\mu{}_\nu -\frac{1}{2}\delta^\mu{}_\nu\delta R~~.  
\label{eq_EinsteinTensorP}
\end{eqnarray}
The perturbed Einstein equation is then
\begin{eqnarray}
\delta R^\mu{}_\nu-\frac{1}{2}\delta^\mu{}_\nu\delta R=8\pi G\delta 
{T^\mu{}_\nu}_{[\mathrm{FL}]}~~.
\label{eq_EinsteinPerturbation} 
\end{eqnarray}
The perfect fluid  energy-momentum tensor has the form
\begin{eqnarray}
 {T^\mu{}_\nu}_{[\mathrm{FL}]}=p_{[\mathrm{FL}]}g^\mu{}_\nu&+(\rho_{[\mathrm{FL}]}+p_{[\mathrm{FL}]})U^\mu U_\nu~~,
   \label{eq_EMTensor}
\end{eqnarray}
where the subscript $[FL]$ denotes the fluid, while $U^\mu = dx^\mu /\sqrt{-ds^2} $ is the four-velocity of the
fluid relative to an observer in the frame in which the Einstein equation (\ref{eq_EinsteinPerturbation}) is solved.  
We define the pressure $p$ and energy density $\rho$ for the ideal fluid which is instantaneously observed by a comoving observer at rest relative to the fluid. We define the velocity fluid $v^i \equiv dx^i/d\tau$, we can treat $v^i$ as the same order perturbation as$\delta\rho_{[\mathrm{FL}]}=\rho_{[\mathrm{FL}]}-\bar{\rho}_{[\mathrm{FL}]}$, $\delta p_{[\mathrm{FL}]}=p_{[\mathrm{FL}]}-\bar{p}_{[\mathrm{FL}]}$, We obtain the energy-momentum tensor for linear order perturbations as follows, 
\begin{eqnarray}
 {T^0{}_0}_{[\mathrm{FL}]}= -(\bar{\rho}_{[\mathrm{FL}]}+\delta \rho_{[\mathrm{FL}]}) ~~,\\
 {T^0{}_i}_{[\mathrm{FL}]}= (\bar{\rho}_{[\mathrm{FL}]}+\bar{p}_{[\mathrm{FL}]})v_i=-T^i{}_0{}_{[\mathrm{FL}]} ~~,\\
 {T^i{}_j}_{[\mathrm{FL}]}= (\bar{p}_{[\mathrm{FL}]}+\delta p_{[\mathrm{FL}]})\delta^i{}_j+\Sigma^i{}_j~ ,~\hspace{.5cm}\Sigma^i
{}_i=0 ~~.
\label{eq:emtensor-elem}
\end{eqnarray}
Since we wish to consider the affect of the PMF, 
we combine  the electromagnetic and perfect fluid energy momentum tensors,
\begin{eqnarray}
T^{\alpha\beta}=
{T^{\alpha\beta}}_{[\mathrm{FL}]}+{T^{\alpha\beta}}_{[\mathrm{EM}]}~~.
\label{eq:all_tensor}
\end{eqnarray}
We use the Boltzmann equations for the flat FRW Universe which Hu and White (1997) derives as follows. 
The moments of the temperature fluctuations
	$\Theta^{(m)}_l$ is
\begin{eqnarray}
\dot{\Theta}^{(m)}_l=k\left[\frac{_0\kappa^m_l}{(2l-1)}\Theta^{(m)}_{l-1}-\frac{_0\kappa^m_{l+1}}{(2l+3)}\Theta^{(m)}_{l+1}\right]\nonumber\\
-\dot{\tau}_c\Theta^{m}_l+S^{m}_l
(l\geqslant m)~~,
\label{eq:boltzmann}
\end{eqnarray}
 where 
\begin{eqnarray}
_s\kappa^{m}_l=\sqrt{\frac{(l^2-m^2)(l^2-s^2)}{l^2}}~~.
\end{eqnarray}
and the source term $S^{(m)}_l$ is 
\begin{eqnarray}
S^{(m)}_l=
\left(
	\begin{array}{ccc}
	\dot{\tau}_c\Theta^{(0)}_0-\dot{\phi} & \dot{\tau}_c v^{(0)}_b+k\psi & \dot{\tau}_c P^{(0)} \\
	0 & \dot{\tau}_c v^{(1)}_b+\dot{V} & \dot{\tau}_c\dot{P}^{(1)} \\
		0 & 0 & \dot{\tau}_cP^{(2)}-\dot{H}
\end{array}
\right)~~.
\label{eq:temperature}
\end{eqnarray}
In equations (\ref{eq:boltzmann}) and (\ref{eq:temperature}) we have introduced the optical depth $\tau_c$ between epoch $\tau$ and the present epoch $\tau_0$.
\begin{equation}
\tau_c(\tau) \equiv \int_\tau^{\tau_0} \dot\tau_c(\tau') d\tau' ~~,
\end{equation}
Also in Eq. (\ref{eq:temperature}) we have introduced the anisotropic nature of Compton scattering as follow,
\begin{eqnarray}
P^{(m)}=\frac{1}{10}[\Theta^{(m)}_2-\sqrt{6}E^{(m)}_2]~~.
\end{eqnarray}
The polarization evolutions are as follows, 
\begin{eqnarray}
\dot{E}^{(m)}_l=
k\left[\frac{_2\kappa^m_l}{(2l-1)}E^{(m)}_{l-1}-\frac{2m}{l(l+1)}B^{(m)}_l-\frac{_2\kappa^m_{l+1}}{(2l+3)}E^{(m)}_{l+1}\right]\nonumber\\
-\dot{\tau}_c[E^{m}_l+\sqrt{6}P^{(m)}\delta_{l,2}](l\geqslant m)~~,
\end{eqnarray}
and
\begin{eqnarray}
\dot{B}^{(m)}_l=
k\left[\frac{_2\kappa^m_l}{(2l-1)}B^{(m)}_{l-1}+\frac{2m}{l(l+1)}E^{(m)}_l-\frac{_2\kappa^m_{l+1}}{(2l+3)}B^{(m)}_{l+1}\right]\nonumber\\
-\dot{\tau}_cB^{m}_l
(l\geqslant m)~~.
\end{eqnarray}
\subsection{Scalar Mode}
The matter fluid is assumed that it is representable as a ideal fluid and neglect the anisotropic pressure perturbations. 
Only adiabatic perturbations is considered and the entropy perturbations is negligible
The linearized perturbation equations which are obtained (Ma \& Bertschinger, 1995; Hu \& White, 1997; Koh \& Lee, 2000) from the Einstein equations up to first order are:
\begin{eqnarray}
	3H^2\psi+3H\dot{\phi}+k^2\phi&=&
		4\pi Ga^2 \delta T^0_0,\label{eq:scalar_Einstein_00}\\
	H\psi+\dot{\phi}&=&
		4\pi G a^2 (\rho+p)v^{(0)}, \label{eq:scalar_Einstein_0j}\\
	\ddot{\phi}
	+H(\dot{\psi}+2\dot{\phi})
	+\left(2\frac{\ddot{a}}{a}-H^2\right)\psi
	+\frac{k^2}{3}(\phi-\psi)&=&
		\frac{4\pi}{3}Ga^2\delta T^i_i, \label{eq:scalar_Einstein_ii}\\
	k^2(\phi-\psi)&=& 
		12\pi Ga^2(\rho+p)\sigma, \label{eq:scalar_Einstein_ij}
\end{eqnarray}
where
\begin{eqnarray}
\rho=\rho_{[\mathrm{FL}]}+\rho_{[\mathrm{EM}]}\label{eq:all_density},\\
p=p_{[\mathrm{FL}]}+p_{[\mathrm{EM}]}\label{eq:all_pressure},
\end{eqnarray}
while $v^{(0)}$ and $\sigma$ are defined as
\begin{eqnarray}
	(\rho+p)v^{(0)}\equiv
		ik^j\delta T^0_j\label{eq:scalar_theta}\\
	(\rho + p)\sigma\equiv
			-\left(\frac{\mathrm{k}_i\cdot\mathrm{k}_i}{k^2}
			-\frac{\delta_{ij}}{3}\right)\Sigma^i_j\label{eq:scalar_sigma}
\end{eqnarray}
and
\begin{eqnarray}
	\Sigma^i{}_j\equiv
		T^i{}_j-\frac{\delta^i{}_jT^k{}_k}{3}
\end{eqnarray}
denotes the traceless component of $T^i{}_j$.
The conservation of energy-momentum is a consequence of the Bianchi identitiy;
 \begin{eqnarray}
T^{\mu\nu}{}_{;\mu}=
	\partial_\mu T^{\mu\nu}
	+\Gamma^\nu{}_{\alpha\beta}T^{\alpha\beta}
	+\Gamma^\alpha{}_{\alpha\beta}T^{\nu\beta}
	=0~~.
\label{eq:conservation_equations}
\end{eqnarray}
Including the energy momentum tensor for the PMF and  writing the magnetic energy density as 
\begin{eqnarray}
	\rho_{[\mathrm{EM}]}=\frac{{B_\lambda}^2}{2\pi}
\label{eq:EM_energy_density},
\end{eqnarray}
Eq.~(\ref{eq:conservation_equations}) leads to the following equations in $k$-space:
\begin{eqnarray}
\dot{\delta}=-(1+w)\left(v^{(0)}-3\dot{\phi}\right)
			-3H\left(\frac{\delta p}{\delta\rho}-w\right)\delta~~,
\label{eq:scalar_contiunuity1}\\
\dot{v^{(0)}}=-H(1-3w)v^{(0)}
			-\frac{\dot{w}}{1+w}v^{(0)}
			+\frac{\delta p}{\delta\rho}\frac{k^2\delta}{1+w}
			-k^2\sigma+k^2\psi~~,
\label{eq:scalar_motion}
\end{eqnarray}
where $w \equiv p/\rho$.
Here, we can cancel $\sigma_{[\mathrm{EM}]}$ in Eq.~(\ref{eq:scalar_motion}) in the linear perturbation limit.  Thus, in the motion and continuity equations for the scalar mode, we can just add the energy density and pressure of the PMF to the general energy density and pressure respectively. From equations (\ref{eq:scalar_contiunuity1}) and (\ref{eq:scalar_motion}) we obtain the same form for the evolution equations of photons and baryons as that deduced in  previous works (Ma \& Bertschinger 1995; Hu \& White 1997).
\begin{eqnarray}
 \dot{\delta}_{\gamma}&=&-\frac{4}{3}v^{(0)}_{\gamma}-4\dot{\phi}~~,\label{eq:photon_delta}\\
 \dot{v}^{(0)}_{\gamma}&=&k^2\left(\frac{1}{4}\delta_{\gamma}-\sigma_{\gamma}\right)
+k^2\psi+an_e\sigma_T(v^{(0)}_{b}-v^{(0)}_{\gamma})~~,\label{eq:photon3}\\
 \dot\delta_b &=& -v^{(0)}_b+3\dot{\phi}~~, \label{eq:baryon_delta}\\
 \dot{v}^{(0)}_b&=&-\frac{\dot{a}}{a}v^{(0)}_b+c^2_sk^2\delta_b+\frac{4\bar{\rho}
 _\gamma}{3\bar{\rho}_b}an_e\sigma_T(v^{(0)}_{\gamma}-v^{(0)}_b)+k^2\psi~~, \label{eq:baryon1} 
\end{eqnarray}
where $n_e$ is the free electron density, $\sigma_T$ is the Thomson
scattering cross section, and $\sigma_{\gamma}$ in the second term on
the right hand side of equation (\ref{eq:photon3}) is the
shear stress of the photon with the PMF. 
In the presence of the PMF, the magnetic pressure should be included in the
acoustic term. Therefore, the sound speed in the second term on the
right hand side of Eq.(\ref{eq:baryon1}) should be
\begin{equation}
 c_s^2=c_{bs}^2 + x_e(\tau)\Pi^{(0)}/4\pi a^4\rho_b, 
\end{equation}
where $x_e(\tau)$ is ionization ratio (we have solved $x_e$ by using the
RECFAST code \cite{recfast}) and $c_{bs}$ is the baryon sound speed without the PMF (see Adams et al 1996). 
\subsection{Vector Mode}
The evolution of the vector potential $V(\tau, \mathbf{k})$ under the influence of the stochastic PMF can be written (Hu and White 1997; Mack et al. 2002) as
\begin{eqnarray}
\dot{V} (\tau, \mathbf{k})+2\frac{\dot{a}}{a}V(\tau, \mathbf{k})
=-\frac{16\pi GL^{(1)} (\tau, \mathbf{k})}{a^2k^2}\nonumber \\
-8\pi Ga^2\frac{p_\gamma\pi_\gamma+p_\nu\pi_\nu}{k}~~,
\label{eq:vectorpoten}
\end{eqnarray}
where the dot denotes a conformal time derivative, $p_i$ and $\pi_i$ are the pressure and the anisotropic stress of the photons ($i=\gamma$) and neutrinos ($i=\nu$).
Since the vector anisotropic stress of the fluid is negligible generally,
it is omitted.

The magnetic field affects the photon-baryon fluid
dynamics via a Lorentz force
term in the baryon Euler equations. Following Hu and White (1997), the Euler equations for neutrino, photon and baryon velocities, $v_\nu$, 
$v_\gamma$, and $v_b$ are written as
\begin{eqnarray}
\dot{v}_{\nu}^{(1)}-\dot{V}=-k\left(\frac{\sqrt{3}}{5}\Theta^{(1)}_{\nu 2}\right)~~,
\label{eq:vneutrino1}\\
\dot{v}_{\gamma}^{(1)}-\dot{V}+\dot{\tau}_c(v^{(1)}_{\gamma}-v^{(1)}_{b})=-k\left(\frac{\sqrt{3}}{5}\Theta^{(1)}_{\gamma 2}\right)~~,
\label{eq:vphoton1}\\
\dot{v}_{b}^{(1)}-\dot{V}+\frac{\dot{a}}{a}(v_{b}^{(1)}-V)-R\dot{\tau}_c(v^{(1)}_{\gamma}-v^{(1)}_{b}) \nonumber \\
=\frac{L^{(1)}(\tau, \mathbf{k})}{a^4(\rho_b+p_b)}~~,
\label{eq:vbaryon1}
\end{eqnarray}
where $R\equiv(\rho_\gamma+p_\gamma)/(\rho_b+p_b)\simeq (4/3)(\rho_\gamma/\rho_b)$ is the inertial density ratio between baryons and photons, while
$\Theta^{(1)}_{\nu 2}$ and $\Theta^{(1)}_{\gamma 2}$ are quadrupole moments of the neutrino and photon angular distributions, respectively. These quantities are proportional to the anisotropic stress tensors. Equations (\ref{eq:vneutrino1})-(\ref{eq:vbaryon1}) denote the vector equations of motion for the cosmic fluid, which arise from the conservation of energy-momentum. 

\section{Temperature Power Spectra}
We have constructed a Boltzmann code for the vector mode by explicitly
expanding CMBFAST (Seljak \& Zaldarriaga 1997).
The temperature transfer function of the CMB today for the scalar and vector mode is
expressed as an integral along the line of sight for $l\ge 2$(Hu and White,1997; Mack et al., 2002)
\begin{eqnarray}
 \frac{\Theta^{(0)}_l(\tau_0, k)}{2l+1}= \int^{\tau_0}_0d\tau e^{-\tau_c}
 	[
		(
	 		\dot{\tau_c}\Theta^{(0)}_0
 			+\dot{\tau_c}\psi+\dot{\psi}-\dot{\phi}
 		)
 		j^{(00)}_l\nonumber\\
 		+\dot{\tau_c}v^{(0)}_bj^{(10)}_l
 		+\dot{\tau_c}P^{(0)}j^{(20)}_l
 	]~~,
\label{eq:Theta_scalar} \\
 \frac{\Theta^{(1)}_l(\tau_0, k)}{2l+1}= \int^{\tau_0}_0d\tau e^{-\tau_c}
 \left\{[\dot{\tau}_c (v^{(1)}_b-V)]D_1\frac{j_l(x)}{x}\right. \nonumber \\
 \left.+\left(\dot{\tau}_c P^{(1)}+\frac{\sqrt{3}}{3}kV\right)D_2\frac{d}{dx}\left(\frac{j_l(x)}{x}\right)\right\}~~,\label{eq:Theta} 
\end{eqnarray}
where the $j_l(x)$ are the spherical Bessel functions with $x=k(\tau_0-\tau)$, $D_1=\sqrt{l(l+1)}/2$, $D_2=\sqrt{3}D_1$, and $P^{(1)}$ is the polarization term (Hu and White 1997).
Using $\Theta_l^{(1)}(\tau_0, k)$, we can compute angular power spectrum of temperature anisotropies:
\begin{eqnarray}
 C_l^{(0)} = \frac{4}{\pi}\int dk k^2 \frac{|\Theta_l^{(0)}(\tau_0,k)|^2}{(2l+1)^2}~~, \\
 C_l^{(1)} = \frac{4}{\pi}\int dk k^2 \frac{|\Theta_l^{(1)}(\tau_0,k)|^2}{(2l+1)^2}~~.
 \end{eqnarray}
This power spectrum of temperature anisotropies is to be  added to that induced
by the scalar density perturbations and compared with observational data.
\section{Result and Discussions}
We discuss the effects of scalar and vector modes of the PMF and also the SZ effect on the CMB in this section.
Note that the current data for higher $l$ are known to be insensitive to the tensor mode which we ignored in the present calculations.
Figures 1 and 2 show the CMB power spectra generated by the PMF.
The effects of the PMF become progressively more important at higher multipoles and eventually dominates over the scalar fluctuations for $l >1000$. 
As a result, the potential discrepancy between the observed and theoretical CMB temperature anisotropies at higher multipoles $l$ is remarkably relaxed by the effects of the presence of a PMF (Yamazaki et al 2005). 
In the following subsections we discuss details on the behavior of the effects of the PMF on the CMB temperature anisotropies.

\subsection{Scalar and Vector modes}
The left hand side of Figure 1 shows the temperature anisotropies of the CMB power spectra which are generated by the PMF, and the right side hand of Figure 1 shows the primary temperature anisotropies calculated in the $\Lambda$ Cold Dark Matter (LCDM) model with and without the effects of the PMF. 
The scalar part of the primary temperature anisotropies consists of several different sources of metric perturbations including photon and neutrino energy density perturbations as well as the perturbation due to the PMF.
Therefore, the scalar mode generated by the PMF, as displayed by the green dash-dotted curves on the left hand side of Fig. 1, is evaluated by subtracting the two calculated CMB power spectra with and without the PMF in the LCDM model.
Note that there is no such complication for the vector part of the primary temperature anisotropies because it arises uniquely from the vector mode of the PMF.

Both scalar and vector modes become larger with increasing field strength $|B_\lambda |$ of the PMF, but the variation of the vector mode is much stronger than the scalar mode (see the right hand side of Fig. 1).
The scalar mode for the PMF gets complicated as is to be discussed later in this subsection. However the net effect is basically caused by two contributions:  First, the increase of the sound speed for additional magnetic pressure to reduce baryonic matter fluctuation amplitude; and second, the changes of the metric and the total energy density which offset the first effect. 
Since the effect of the PMF for the vector mode is just an increase of the perturbation by the Lorentz force, the vector-to-scalar mode ratio increases as the strength of the PMF $|B_\lambda |$ increases.

The PMF affects both the energy density and pressure in the scalar mode. The influence on the energy density is the same as that on the baryon density. 
The magnetic pressure, however, increases the sound speed of the fluid as discussed below Eqs.(\ref{eq:baryon_delta}) and (\ref{eq:baryon1}).  This produces interesting  effects on the CMB (Adams et al 1996; Tsagas and Maartens 2000; Yamazaki et al 2006). 
In order to understand the effects of the change of the sound speed, $c_s$, boosted by the PMF let us solve Eqs.(\ref{eq:baryon_delta}) and (\ref{eq:baryon1}) in the WKB  approximation 
\begin{eqnarray}
 \delta_b &\propto & 
 - \frac{1}{\sqrt{c_s(\tau) a(\tau)}}
 \exp\left(-i\int c_s kd\tau \right). \label{eq:sound2}
\end{eqnarray}
This solution clearly shows two different influences of the change of $c_s$ on the baryonic matter fluctuation: 
First, the increase of the sound speed due to the magnetic pressure results in a decrease in the amplitude of the evolution of $\delta_b$ because of the prefactor of Eq.(\ref{eq:sound2}). 
This means that in the presence of a magnetic field, the plasma pressure increases by the repulsion of lines of magnetic force.  Thus,  gravitational collapse of the matter is delayed. 
Second, increasing the sound velocity makes the frequency of the baryonic fluctuation $\delta_b$ larger because of the exponent of Eq.(\ref{eq:sound2}). 
This effect of the PMF on density field makes different kinds of amplitude boosts which depend on $k$ (Yamazaki et al. 2006).
The PMF shifts the acoustic oscillation for the baryon fluid  either up or down depending on the combination of the sound speed $c_s$ and wavenumber $k$. 
Note, however, that since the amplitude of the PMF is so small, the difference between the shifted and non-shifted period of the acoustic oscillations essentially only depends on the wavenumber $k$. This effect is strongly constrained by the observed data as displayed on the right hand side of Figure 1.

When one compares the scalar and vector modes, the scalar mode makes only a minor contribution to the CMB except at lower amplitudes. However, since the magnetic field pressure delays the gravitational collapse of matter, its effect on the matter power spectrum for large scale structure (LSS) is very interesting. 
Since in the present article we are only interested in the small scale structure associated with larger multipoles $l$, further details on this point will be discussed elsewhere (Yamazaki et al. 2006).

\subsection{Dependence on $B_\lambda$}
The CMB power spectrum generated by a cosmological magnetic field is generally characterized by a broad peak at $l\simeq$1000$\sim$3000 for a reasonable range of -2$<n_B\lesssim$0 as shown in Fig.~2. 
We display five cases of the CMB power spectra for $B_\lambda$=1, 2, 4, 8, and 10 $n$G in Fig.1.
As mentioned above, for lower $l$ and $B_\lambda$,  the scalar mode dominates (bottom three cases for $B_\lambda = $1, 2 and 4 $n$G of the left hand side of Fig.1). For higher $l$ and $B_\lambda$, the vector mode dominates.
The amplitude is shown to increase with increasing magnetic field strength $B_\lambda$ for $l > 1000$, where the  vector mode dominates. Some lengthy mathematical derivation of the power spectrum of fluctuations leads to $C_l \propto B^4_\lambda$ for $n_B<$-3/2, and $C_l \propto B_\lambda^{{14}/{(n_B+5)}}$ for $n_B>$-3/2 for higher multipoles $l$ (Fig.~2).

\subsection{Dependence on $n_B$}
Fig.~\ref{fig2} shows the $n_B$ dependence of the CMB.
For $n_B<$-3/2, the second term in Eq.(\ref{eq:Pi}) dominates.  This $k$-dependence causes a strong angular dependence for the vector mode perturbation. This results in changing the peak position for different multipoles $l$ depending upon the power-law spectral index $n_B$.
On the other hand, peaks remain at nearly the same multipole $l\sim$ 2000 independently of $n_B$ for $n_B>$-3/2. 
This is because the vector mode perturbation from the  magnetic field tends to be that of white noise when the first term in Eq.(\ref{eq:Pi}) dominates. 

\section{Constraint on the PMF in a MCMC Analysis}
To set limits on the PMF we have evaluated likelihood functions for fits to the  WMAP, ACABAR, and CBI data over a wide range of the parameters, $B_\lambda$
and $n_B$, for a stochastic PMF along with the usual six cosmological parameters, $h, \Omega_b h^2,\Omega_m h^2, n_s, A_s$, 
and $\tau _c$. 
The quantity $h$ is the Hubble constant in units of 100 km s$^{-1}$ Mpc$^{-1}$.
The quantities  $\Omega_b h^2$ and $\Omega_m h^2$ are
the usual present baryon and cold dark matter closure parameters.  The quantities $n_s$ and $A_s$ are the
spectral index and the amplitude of the primordial scalar fluctuation, respectively.  The parameter
 $\tau _c$ is the optical depth. 
 We adopt a flat cosmology. 
 To explore this parameter space, we have employed
 the Markov chain technique (Lewis 2002).
We also take account of the SZ effect in our analysis. For that, we
adopt a fixed prior of   $\sigma_8=0.9 $ (Spergel
et at. 2003; Komatsu and Seljak 2002).
\subsection{Degeneracy}

In our likelihood analysis of the magnetic field parameters, we continued the MCMC algorithm until the  cosmological parameters well converged to the values listed in Table 1.
The inferred parameter values are consistent with those deduced in other analyses (Spergel et al. 2002; Mason et al., 2003; Kuo et al., 2004).
We did not find obvious degeneracies of the magnetic field parameters with other cosmological parameters.
We understand this for the following reasons. 
The primordial magnetic field is constrained (Jedamzik et al. 1998;  Mack et al. 2002) by the cutoff scale for the  damping of Alfven waves as indicated by Eq.~(\ref{eq:kd}).
Since this cutoff scale is small compared with the multipoles for the currently available data at $l \ll$ 3000,  the manifestation of this effect is remarkably seen as a monotonically increasing contribution from both scalar and vector perturbations in the multipole region higher than $l\sim$500 (Yamazaki et al. 2005a).
This sensitivity of the CMB power spectrum to the primordial magnetic field differs completely from those to the other cosmological parameters, which helps resolve the degeneracies among them.
  There is, however, a strong degeneracy between the magnetic field strength $|B_\lambda|$ and the power spectral index $n_B$
  because those parameters affect the amplitude simultaneously [see Eqs.~(\ref{eq:Pi}) and (\ref{eq:Scalar_Pi})].

\subsection{Limits on the PMF}

 In Fig.~\ref{fig6} we show results of our MCMC analysis using  the WMAP , ACBAR and CBI data in the two parameter plane $|\mathbf{B}_{\lambda}|$ vs. $n_B$.
The 1$\sigma$(68\%)
and 2$\sigma$(95.4\%)
 C.L.~ excluded regions are bounded above by the thick curves as shown. 
 We could not find a lower boundary of the allowed region at the 1$\sigma$ or 2$\sigma$  confidence level. 
Note, however,  that we find a very shallow minimum with a reduced $\chi^2 \simeq 1.08$.
From Fig.~\ref{fig6} we obtain an upper limit to the strength of the PMF at the 1$\sigma$(95.4\%) C.L.~of
\begin{eqnarray}
|B_\lambda|\ \lesssim 7.7\ \mathrm{nG}\ \mathrm{at\ 1Mpc}~~.
\end{eqnarray}
This upper limit is particularly robust as we have considered all effects on the CMB anisotropies, i.e. the effect of the ionization ratio, the SZ effect, and the effects from both the scalar and vector modes on the magnetic field, in the present estimate of $|\mathbf{B}_\lambda|$ and $n_B$.
\section{PMF Generation and Evolution}
In this section we discuss a multiple generation and evolution scenario of the cosmological primordial magnetic field that is motivated by the results of the present study.

To begin with we adopt the following three constraint conditions:

\begin{enumerate}
\item A PMF strength $|\vt{B}_\lambda| \lesssim$7.7nG(1$\sigma$) at 1Mpc as deduced above by applying  the MCMC method to the  WMAP, ACBAR, and CBI data.

\item The magnetic field strength in galaxy clusters is $0.1\mu$G $<\ |B_{CG}|\ <\ 1\mu$ G (Clarke et al. 2001; Xu et al.~ 2006).   Hence, if the isotropic collapse is the  only process which amplifies the  magnetic field strength, the lower limit to the PMF is $\sim 1-10$ nG for the PLSS.

\item The gravity wave constraint on the PMF from Caprini and Durrer (2002)\footnote{$\lambda$ in Caprini and Durer (2004) is for a scale of 0.1 Mpc, however our $\lambda$ is for 1Mpc. Thus, inclinations of lines in Figs. \ref{fig6} and \ref{fig7} are smaller than Caprini and Durer (2004).}.
The big-bang nucleosynthesis of the light elements depends on a balance between the particle production rates and the expansion rate of the universe. 
Since the energy density of the gravity waves $\rho_{\mathrm{GW}}$
contributes to the total energy density.  However,  $\rho_{\mathrm{GW}}$ is constrained so that  the expansion rate of the universe does not spoil the agreement between the theoretical and observed light element abundance constraints for 
deuterium, $^3$He, $^4$He, and $^7$Li (Maggiore, 2000). 
\end{enumerate}

Figures~\ref{fig6} and \ref{fig7} summarize the various  constraints on the PMF by these three conditions. 
The region bounded by the  upper red-solid curve is constrained by condition (\roman{one}) as indicated, the green-dash-dotted horizontal line corresponds to  the lower limit to the PMF from condition (\roman{two}), and the Black-solid, sky-blue-dotted, and pink-dashed lines, respectively, are the upper limit of the produced PMF from big-bang nucleosynthesis, the electroweak transition, and the inflation epoch.
Figure. \ref{fig7} shows the allowed or excluded regions according to these multiple constraints depending upon  when the PMF was generated:

(A). Nucleosynthesis: I+II+III region
\begin{eqnarray}
1.0\ \mathrm{nG}\ \lesssim\ |B_\lambda|\ \lesssim\ 4.7\mathrm{nG} \nonumber\\
-3\ \lesssim\ n_B\ \lesssim\ -2.40 \nonumber
\end{eqnarray}

(B). Electroweak transition: II+III region
\begin{eqnarray}
1.0\ \mathrm{nG}\ \lesssim\ |B_\lambda|\ \lesssim\ 4.7\mathrm{nG} \nonumber\\
-3\ \lesssim\ n_B\ \lesssim\ -2.43 \nonumber
\end{eqnarray}

(C). Inflation: III region
\begin{eqnarray}
1.0\ \mathrm{nG}\ \lesssim\ |B_\lambda|\ \lesssim\ 4.7\mathrm{nG} \nonumber\\
-3\ \lesssim\ n_B\ \lesssim\ -2.76 \nonumber
\end{eqnarray}

Obviously,  the upper limits on both $|B_\lambda|$ and $n_B$ become more stringent if the PMF is produced during an earlier epoch.  
Moreover, the  limits we deduce are the strongest constraints on the PMF that have  yet been determines. However, we caution that the evolution of the generation of the PMF during the LSS epoch is not well understood. Thus, if there are other effective physical processes for the generation and evolution of the PMF during the formation of LSS, our lower limit of the PMF parameters may decrease.
 In order to constrain the PMF parameters accurately, we should study the PMF not before but after the PLSS.

\section{Conclusion}
For the first time we have studied  scalar mode effects of the PMF on the CMB.
We have confirmed numerically without approximation that the excess power in the  CMB at higher $l$ can be explained by the existence of a PMF.
For the first time a likelihood analysis utilizing the  WMAP, ACBAR and CBI data with a MCMC method has been applied to constrain the upper limit on the strength of the PMF to be 
\begin{eqnarray}
|B_\lambda| < 7.7nG ~~.\nonumber
\end{eqnarray}
We have also considered three conditions on the generation and evolution of the cosmological PMF:
1) our result;
2) the lower limit of the PMF from the magnetic field of galaxy clusters;
 and 3) the constraint on the PMF from gravity waves. 
 COmbining these, we find the following concordance region for the  the PMF parameters;
\begin{eqnarray}
1~\mathrm{nG}< |B_\lambda| < 4.7~ \mathrm{nG}~~,\ \ -3.0< n_B < -2.4~~.\nonumber
\end{eqnarray}

The PMF also affects the formation of  large scale structure.  For example, magnetic pressure delays the gravitational collapse. It is thus very important to constrain the PMF as precisely as possible. 
If we combine our study and future plans to observe the CMB anisotropies and polarizations for higher multipoles $l$, e.g.~via the {\it Planck Surveyor}, we will be able to constrain the PMF more accurately, and explain the evolution and generation of the magnetic field on galaxy cluster scales along with  the formation of the LSS.

\acknowledgments{ We acknowledge Drs. K. Saigo, H. Hanayama, M. Higa, R. Nakamura, K. Umezu, H. Ohno, K. Takahashi, M. Oguri, Prof. M. Yahiro for their valuable discussions. This work has been supported in part by Grants-in-Aid for Scientific
Research (13640313, 14540271) and for Specially Promoted Research (13002001)
of the Ministry of Education, Science, Sports and Culture of Japan, and the
Mitsubishi Foundation. K. I. also acknowledges the support by Grant-in-Aid for
JSPS Fellows.   Work at the University of Notre Dame (G.J.M.) supported
by the U.S. Department of Energy under 
Nuclear Theory Grant DE-FG02-95-ER40934.

}

\appendix
\section{Tight Coupling Approximation}
Since the Thomson opacity is larger before recombination, photons and baryons are tightly coupled. If the Thomson drag terms in Eq.(\ref{eq:photon3}), (\ref{eq:baryon1}), (\ref{eq:vphoton1}), and (\ref{eq:vbaryon1}) are too large, it is difficult to solve these equations.
Therefore, we here derive approximation equations in this limit. 
The tight coupling approximation for the scalar mode was introduced by Ma and Rertschinger (1997). Here, therefore, we only need to introduce the vector mode in this appendix. 

Using equations (\ref{eq:vphoton1}) and (\ref{eq:vbaryon1}), we obtain the following equations for the photons and baryons in the tight-coupling approximation.
\begin{eqnarray}
\dot{v}^{(1)}_b=\frac{1}{1+R}\left(-\frac{\dot{a}}{a}v^{(1)}_b-R(\dot{v}^{(1)}_\gamma-\dot{v}^{(1)}_b)-\frac{\sqrt{3}}{5}kR\Theta^{(1)}_{\gamma 2}+\frac{L^{(1)}}{a^4(\rho_b+p_b)}\right)~~,
\label{eq:tcp1}
\end{eqnarray}

\begin{eqnarray}
v^{(1)}_b-v^{(1)}_\gamma
=\frac{1}{\dot{\tau}_c}\left(\dot{v}^{(1)}_\gamma+\frac{\sqrt{3}}{5}k\Theta^{(1)}_{\gamma 2}\right)~~.
\label{eq:tcp2}
\end{eqnarray}
Writing $\dot{v}^{(1)}_\gamma$ as $\dot{v}^{(1)}_b+(\dot{v}^{(1)}_\gamma-\dot{v}^{(1)}_b)$, we can rewrite (\ref{eq:tcp2})
\begin{eqnarray}
v^{(1)}_b-v^{(1)}_\gamma=\frac{1}{\dot{\tau}_c}\left(\dot{v}^{(1)}_b+(\dot{v}^{(1)}_\gamma-\dot{v}^{(1)}_b)+\frac{\sqrt{3}}{5}k\Theta^{(1)}_{\gamma 2}\right)~~,
\end{eqnarray}
 and using equations (\ref{eq:tcp1}) and (\ref{eq:tcp2}), we get
\begin{eqnarray}
v^{(1)}_b-v^{(1)}_\gamma=\frac{1}{(1+R)\dot{\tau}_c}\left(-\frac{\dot{a}}{a}v^{(1)}_b+(\dot{v}^{(1)}_\gamma-\dot{v}^{(1)}_b)+\frac{\sqrt{3}}{5}k\Theta^{(1)}_{\gamma 2}+\frac{L^{(1)}}{a^4(\rho_b+p_b)}\right)~~.\nonumber \\ \label{eq:tcp4}
\end{eqnarray}
Differentiating Eq.~(\ref{eq:tcp4}) and ignoring terms of more than second order in $\dot{\tau}_c$, we get  
\begin{eqnarray}
\dot{v^{(1)}_b}-\dot{v^{(1)}_\gamma}&=&
\frac{\dot{a}}{a}\frac{2R}{(1+R)}(v^{(1)}_b-v^{(1)}_\gamma)\nonumber\\
&+&\frac{1}{(1+R)\dot{\tau}_c}\left[-\frac{\ddot{a}}{a}v^{(1)}_b\right]
\nonumber\\
&+&O(\dot{\tau}^{-2}_c)~~.
\end{eqnarray}
Substituting Eq.~(\ref{eq:tcp4}) into Eq.~(\ref{eq:tcp1}) yields  
\begin{eqnarray}
\dot{v}^{(1)}_\gamma=\frac{1}{R}\left[-\dot{v}^{(1)}_b-\frac{\dot{a}}{a}v^{(1)}_b-\frac{\sqrt{3}}{5}kR\Theta^{(1)}_{\gamma 2}+\frac{L^{(1)}}{a^4(\rho_b+p_b)}\right]~~.
\end{eqnarray}


\clearpage

\begin{deluxetable}{llll}
\tablecolumns{4}
\tablewidth{0pc}
\tablecaption{Calculated LCDM model parameters which best fit the
WMAP data with the effect of primordial magnetic field taken into account.
}
\tablehead{
Parameter & Mean and 68\% C.L. Errors &  95\% C.L. Errors
}
\startdata
$\Omega_b h^2$ & $0.0231_{-0.0010}^{+0.0011}$ & $_{-0.0020}^{+0.0021}$ \\
$\Omega_{CDM} h^2$ & $0.120_{-0.011}^{+0.011}$ & $_{-0.018}^{+0.024}$ \\
$n_s$ & $0.976_{-0.026}^{+0.027}$ & $_{-0.046}^{+0.054}$ \\
$A_s$ & $0.903_{-0.129}^{+0.131}$ & $_{-0.194}^{+0.236}$ \\
$\tau_c$ & $0.126_{-0.039}^{+0.034}$ & $_{-0.10}^{+0.11}$ \\
$h$ & $0.727_{-0.037}^{+0.037}$ & $_{-0.070}^{+0.074}$ \\
\enddata
\tablecomments{
Calculated corresponding cosmic expansion ages are $t_0$/Gyr = $13.35_{-0.21}^{+0.21}$(68\% C.L.) and $13.35_{-0.41}^{+0.41}$(95\% C.L.).}
\end{deluxetable}
\clearpage
\begin{figure}[t]
\epsscale{0.65}
\plotone{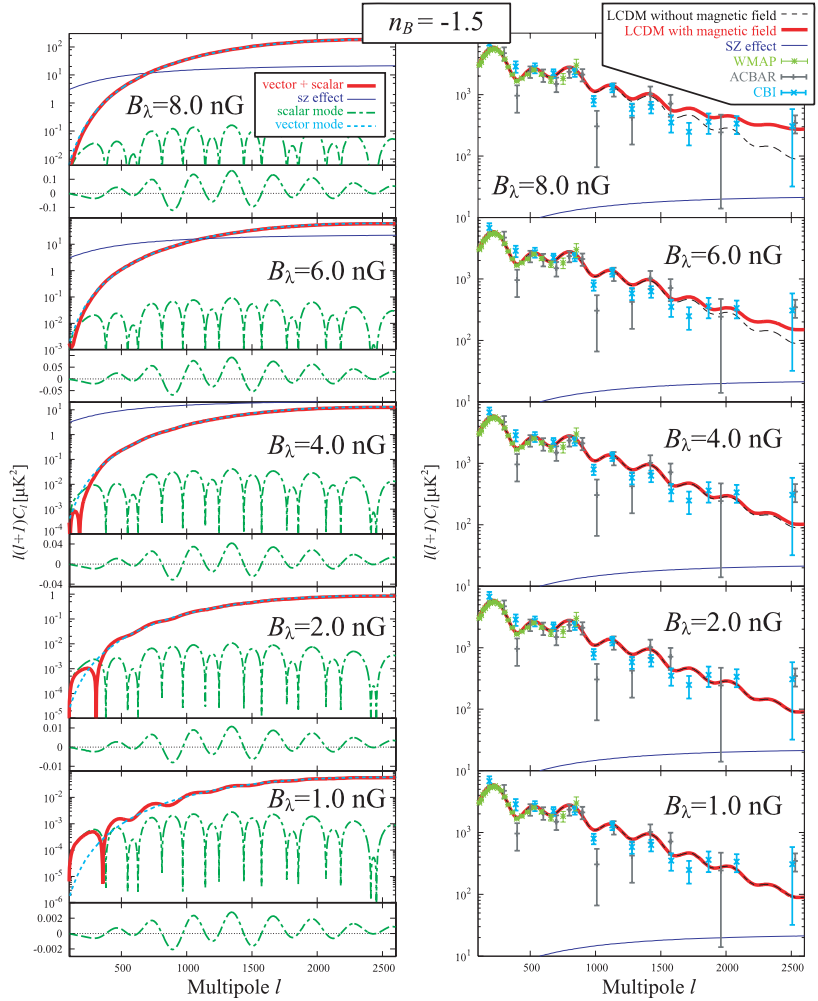}
\caption{Effects of the PMF strength $|B_\lambda|$ on the  CMB. The five figures on the left hand side show the CMB temperature anisotropies from the PMF.  The red bold curves are the temperature anisotropies from vector and scalar modes of the PMF.  The green dashed-dotted curves are for the scalar mode, the skyblue doted curves are for the vector mode, and the purple thin curves show the SZ effect. The upper parts of the figures show the combined  effects on the CMB.  The lower parts show the effects of the  scalar mode obtained by subtracting  the LCDM model without the PMF from the LCDM model with the PMF. 
The five figures on the right hand side show the primary temperature anisotropies of the CMB with and without the PMF. The black short-dashed curves show the LCDM model without the PMF.  The red bold curves are the LCDM model with the PMF.  The purple thin curves are the SZ effect.  The dots with lightgreeen, gray, and skyblue error bars are the WMAP, ACBAR and CBI data, respectivitly.
On all figures, the power spectral index of the PMF is set to $n_B=$-1.5.  Also,  the strengths of the PMF on both sides of this figure are set to be $|B_\lambda| =$ 8.0, 6.0, 4.0, 2.0, and 1.0 nG from top to bottom as labeled.
\label{fig1}}
\end{figure}

\begin{figure}[t]
\epsscale{0.8}
\plotone{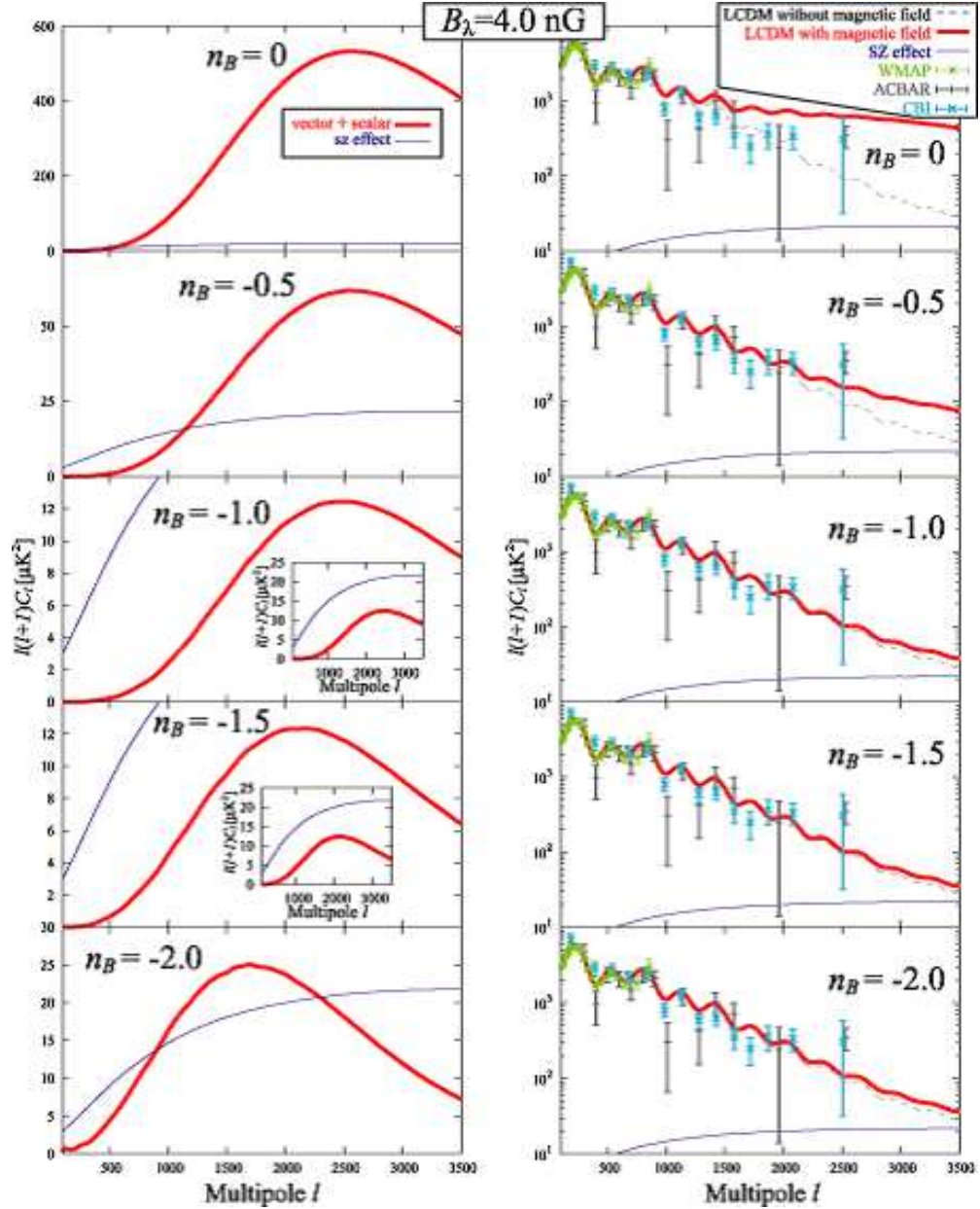}
\caption{Effects  power spectral index $n_B$ of the PMF on the CMB. The five figures on the left hand side show the temperature anisotropies of the CMB from the PMF, and the red bold curves are the temperature anisotropies from the vector and scalar modes of the PMF. The purple thin curves show the SZ effect adopted here. 
The five figures on right hand side show the primary temperature anisotropies of the CMB with and without the PMF. The black short-dashed curves are the LCDM model without the PMF.  The red bold curves are the LCDM model with the PMF.  The purple thin curves are the SZ effect.  The dots with lightgreeen, gray, and skyblue error bars are WMAP, ACBAR and CBI data, respectivitly.
On all figures, the strength of the PMF is set to be $|B_\lambda|=$ 4.0 nG.  Also the power spectral index of the PMF on both sides of this figure are set to $n_B =$  0, -0.5, -1.0, -1.5, and -2.0 from top to bottom as labeled.
\label{fig2}}
\end{figure}

\begin{figure}[t]
\epsscale{0.8}
\plotone{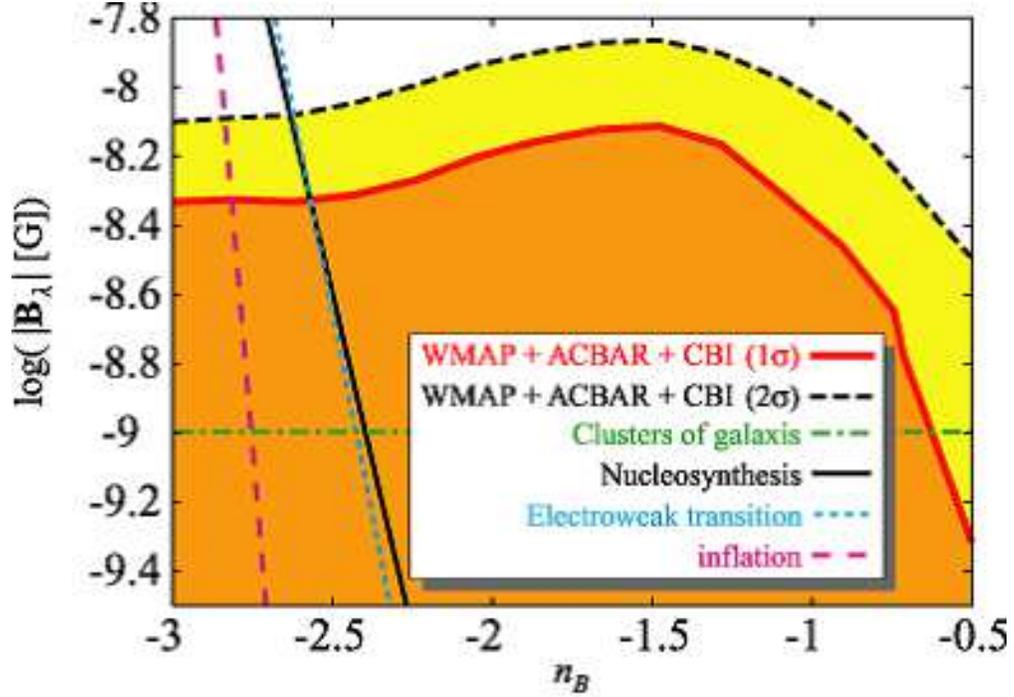}
\caption{Results of the MCMC method constrained by the  WMAP, ACBAR, and CBI data. 
Excluded and allowed regions at the $1\sigma$(68\%) C.L. and $2\sigma$(95.4\%) C.L.~ are shown
 in the two parameter plane of $|\mathbf{B}_\lambda|$ vs. $n_B$, where  $|\mathbf{B}_\lambda|$ is the primordial magnetic field strength and $n_B$ is the power-law spectral index. Solid and dotted curves are for $\Delta\chi^2=$2.3 and 6.17, respectively. The green dash-dotted horizontal line displays the lower limit of the field strength $B_\lambda=1$ nG deduced for  the galaxy cluster scale at the PLSS (Clarke, et al 2001; Xu, et al 2006). The black-solid, skyblue-dotted, and pink-dashed lines are the upper limit of the produced PMF at the big-bang nucleosynthesis, the electroweak transiton, and the inflation epochs, respectively  (Caprini and Durrer 2002).
 \label{fig6}}
\end{figure}

\begin{figure}[t]
\epsscale{0.8}
\plotone{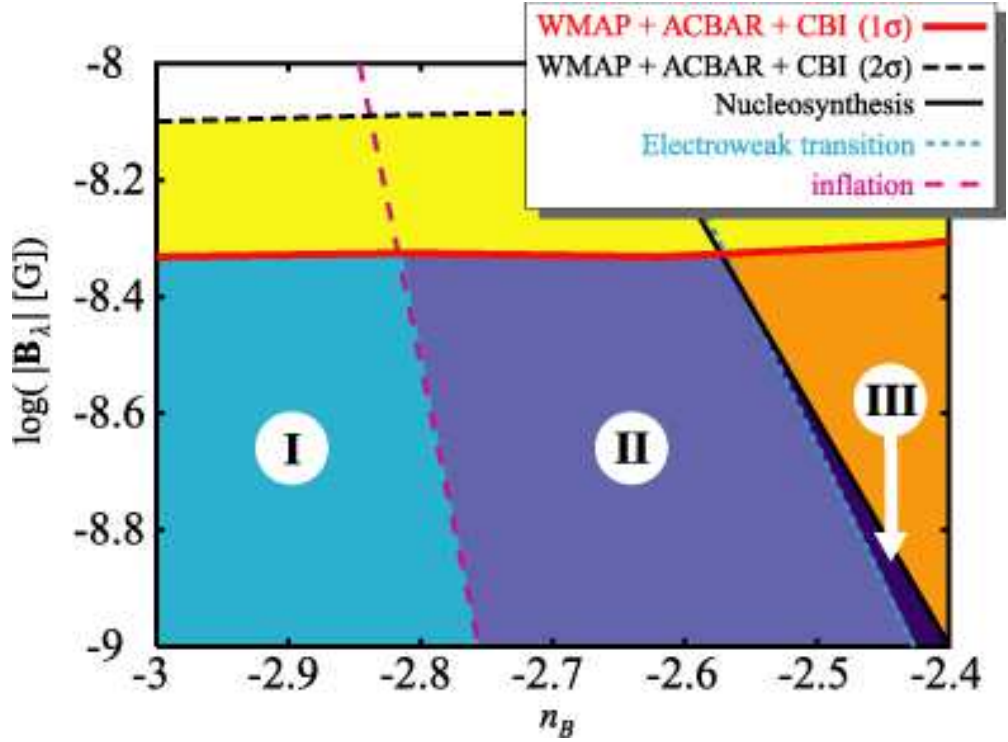}
\caption{Excluded and allowed regions at the $1\sigma$(68\%) C.L. and $2\sigma$(95.4\%) C.L.
in the two parameter plane $|\mathbf{B}_\lambda|$ vs. $n_B$ obtained by the MCMC method with the WMAP, ACBAR, and CBI data. $|\mathbf{B}_\lambda|$ is the primordial magnetic field strength and $n_B$ is the power-law spectral index. The solid and dotted curves are for $\Delta\chi^2=$2.3 and 6.17, respectively. The  black-solid, skyblue-dotted, and pink-dashed lines are the upper limit of the produced PMF at the big-bang nucleosynthesis, the electroweak transition, and the inflation epochs, respectively (Caprini and Durrer 2002). If the PMF is produced at the epochs of big-bang nucleosynthesis, the electroweak transition, or inflation,  respectively, the region of \Roman{one}+\Roman{two}+\Roman{three}, II+III, or III are allowed by these constraints on the PMF for the galaxy cluster scale at the PLSS and the MCMC method with WMAP, ACBAR, and CBI data.
  \label{fig7}}
\end{figure}
\end{document}